# A 'checkerboard' electronic crystal state in lightly hole-doped $Ca_{2-x}Na_xCuO_2Cl_2$

T. Hanaguri[1,2], C. Lupien[3], Y. Kohsaka[4], D.-H. Lee[5,6], M. Azuma[2,7], M. Takano[7], H. Takagi[1,2,4] & J. C. Davis[3]

[1]*Magnetic Materials Laboratory, RIKEN (Institute of Physical and Chemical Research), Wako 351-0198, Japan*

[2]*Japan Science and Technology Agency, Kawaguchi 332-0012, Japan*

[3]*LASSP, Department of Physics, Cornell University, Ithaca, New York 14853 USA*

[4]*Department of Advanced Materials Science, University of Tokyo, Kashiwa 277-8651, Japan*

[5]*Department of Physics, University of California, and* [6]*Material Sciences Division, Lawrence Berkeley National Laboratory, Berkeley, California 94720, USA*

[7]*Institute for Chemical Research, Kyoto University, Uji 601-0011, Japan*

**The phase diagram of hole-doped copper oxides shows four different electronic phases existing at zero temperature. Familiar among these are the Mott insulator, high-transition-temperature superconductor and metallic phases. A fourth phase, of unknown identity, occurs at light doping along the zero-temperature bound of the 'pseudogap' regime[1]. This regime is rich in peculiar electronic phenomena[1], prompting numerous proposals that it contains some form of hidden electronic order. Here we present low-temperature electronic structure imaging studies of a lightly hole-doped copper oxide: $Ca_{2-x}Na_xCuO_2Cl_2$. Tunnelling spectroscopy (at energies $|E|>100$ meV) reveals electron extraction probabilities greatly exceeding those for injection, as anticipated for a doped Mott insulator. However, for $|E|<100$ meV, the spectrum exhibits a V-shaped energy gap centred on $E=0$. States within this gap undergo intense spatial modulations, with the spatial correlations of a four $CuO_2$-unit-cell square 'checkerboard', independent of energy. Intricate atomic-scale electronic structure variations also exist within the checkerboard. These data are consistent with an unanticipated crystalline electronic state, possibly the hidden electronic order, existing in the zero-temperature pseudogap regime of $Ca_{2-x}Na_xCuO_2Cl_2$.**

Scanning tunnelling microscopy (STM) has recently emerged as a suitable technique to search for 'hidden' electronic order in the copper oxides. $Bi_2Sr_2CaCu_2O_{8+\delta}$ (Bi-2212) studies where high-transition-temperature ($T_c$) superconductivity (HTSC) was suppressed to reveal the pseudogap have been especially fruitful. The prototypical study yielded a pseudogap-like conductance spectrum (V-shaped without coherence peaks) associated with a 'checkerboard' approximately four $CuO_2$ unit cells square of local-density-of-states (LDOS) modulations surrounding vortex cores[2]. Similar phenomena were discovered throughout the sample above $T_c$ (ref. 3), and within strongly underdoped nano-regions exhibiting pseudogap-like spectra[4] (see Fig. 1c). Underdoped Bi-2212 therefore shows a tendency towards checkerboard electronic modulations when HTSC is suppressed. However, it is unclear whether these checkerboard modulations in Bi-2212 represent a true electronic phase, because they exhibit[2–4] (1) a variety of doping-dependent incommensurate wavevectors, (2) very weak intensities, and (3) short (~8 nm) correlation lengths within the nanoscale disorder[4–7]. Furthermore, their atomic-scale spatial and energetic structures are unknown, presumably because of disorder[4–7] and/or thermal energy smearing $\Delta E \approx 3.5 k_B T \approx 30$ meV at $T \approx 100$ K (ref. 3).

To search for electronic order hidden in the pseudogap while avoiding these uncertainties, we decided to study a simpler and less disordered copper oxide at lower doping and temperature. We chose $Ca_{2-x}Na_xCuO_2Cl_2$ (Na-CCOC), a material whose parent compound $Ca_2CuO_2Cl_2$ is a canonical Mott insulator[8]. Within its undistorted tetragonal crystal structure (Fig. 1b), all the $CuO_2$ planes are crystallographically identical. Sodium substitution for Ca destroys the Mott insulator state, producing first a nodal metal[9] in the zero-temperature pseudogap (ZTPG) regime, and eventually HTSC for $x \geq 0.10$ (refs 10, 11). Crucially, Na-CCOC is easily cleavable between CaCl layers to reveal an excellent surface. Initial STM studies showed clean, flat CaCl surfaces (with traces of nanoscale electronic self-organisation) which exhibit a V-shaped spectral gap for $|E| < \sim 100$ meV (ref. 12).

Our studies used $Ca_{2-x}Na_xCuO_2Cl_2$ samples with Na concentrations $x$=0.08, 0.10, and 0.12 and bulk $T_c \approx 0$, 15 and 20 K respectively. Atomically flat surfaces are obtained by cleaving below 20 K in the cryogenic ultrahigh vacuum of a dilution refrigerator. Figure 1d shows a typical topographic image of the CaCl plane with inset showing the quality of atomic resolution achieved throughout. These surfaces exhibit a perfect square lattice, without discernible crystal distortion or surface reconstruction, and with lattice constant $a_0$ in agreement with X-ray diffraction (3.85 Å).

To image the electronic states in Na-CCOC, we use spatial- and energy-resolved differential tunnelling conductance, $g(\mathbf{r}, E=eV_s)$; measurements from STM. For a strongly correlated system such as a lightly doped Mott insulator, $g(\mathbf{r}, E)$ is proportional to the momentum-space integrated spectral function at $\mathbf{r}$ (ref. 13), rather than $LDOS(\mathbf{r}, E)$.

Nevertheless, $g(\mathbf{r}, E)$ measurements remain a powerful tool for determining atomic-scale spatial rearrangements of electronic structure.

The properties of $g(\mathbf{r}, E)$ should be determined primarily by states in the $CuO_2$ plane because the CaCl layers are strongly insulating. In support of this, we find that missing surface atoms (arrow in Fig. 1d) do not affect $g(\mathbf{r}, E)$. A typical spatially averaged spectrum $\langle g(\mathbf{r}, E)\rangle$ for $x$=0.12 is shown in black in Fig. 1c. The high energy conductance for electron extraction is ~5 times greater than that for injection and this ratio grows rapidly with falling doping (Supplementary Fig. 1). Such strong bias asymmetries in conductance have long been anticipated. This is because, in a lightly hole-doped Mott insulator, the reservoir of states from which electrons can be extracted at negative sample bias is determined by 1−$p$, while that of hole-states into which electrons can be injected at positive sample bias is determined by $p$, where $p$ is the number of holes per $CuO_2$.

For $|E|$<100 meV, a slightly skewed V-shaped gap, reaching very low conductance at $E$=0, is seen in all Na-CCOC spectra (vertical lines in Fig. 1c). Although whether this gap represents precisely the same phenomena as the Bi-2212 pseudogap above $T_c$ (ref. 1) is unknown, it has (modulo thermal smearing effects) very similar basic characteristics: a V-shaped energy gap spanning approximately ±100 meV with minimum at $E$=0. Furthermore, tunnelling spectra from the most underdoped nano-regions of $p$=0.11±0.01 Bi-2212 (blue symbols in Fig. 1c) are almost identical[4]. It seems reasonable that very similar spectra at similar doping in very different lightly doped copper oxides represent the same electronic state. For these reasons, we provisionally designate this spectrum as characteristic of whatever electronic state exists in the ZTPG regime.

We next focus on imaging states within this pseudogap. In Fig. 1e, we show $g(\mathbf{r}, E$=24 meV) measured in the field of view of Fig. 1d. It is reasonably typical of all $g(\mathbf{r}, E)$ obtained for 0.08≤$x$≤0.12 for 10 meV<$|E|$<100 meV. The most striking fact, and a central result of this paper, is that a clear checkerboard pattern of intense conductance modulations, with primary periodicity $4a_0$, but also complex internal structure at the atomic-scale, is observed.

In Fig. 2a–c we show $g(\mathbf{r}, E)$ measured in the same field of view as Fig. 1d at energies 8, 24 and 48 meV. These and all other subgap $g(\mathbf{r}, E)$ are quite similar. The periodicities of modulations in $g(\mathbf{r}, E)$ are examined via their Fourier transforms $g(\mathbf{q}, E)$, as shown in Fig. 2d–f. These $g(\mathbf{q}, E)$ do not break symmetry under 90° rotations, an observation also noted at $|E|$<100 meV for all dopings. They also reveal three primary sets of **q**-vectors contributing to the subgap $g(\mathbf{q}, E)$: $\mathbf{q}=(\pm 1, 0); (0, \pm 1)2\pi/a_0$ from the lattice, $\mathbf{q}=(\pm 1/4, 0); (0, \pm 1/4)2\pi/a_0$ from the commensurate $4a_0 \times 4a_0$ modulation, and an unanticipated $\mathbf{q}=(\pm 3/4, 0); (0, \pm 3/4)2\pi/a_0$ modulation, which is also intense. Plots of Fourier intensity along the two orthogonal lines $(2\pi, 0)$ and $(0, 2\pi)$ in Fig. 2g–i show these effects directly. Significantly, these sets of **q**-vectors do not disperse with increasing energy and can also be detected as modulations in topographic images (Supplementary Fig. 2). All these observations are consistent with a crystalline electronic state.

To investigate spatial characteristics, we analyse autocorrelation images $A(\mathbf{R}, E)=\langle g(\mathbf{r}, E) g(\mathbf{r}+\mathbf{R}, E)\rangle$. For example, $A(\mathbf{R}, E=24\text{ meV})$ is shown in Fig. 3a. It clearly exhibits a basic $4a_0 \times 4a_0$ spatial modulation with a correlation length of $\sim 10a_0$. For brevity, we refer to the $4a_0 \times 4a_0$ unit cell as a 'tile'. Similar $A(\mathbf{R}, E)$ are found for all $|E|<100$ meV. The $A(\mathbf{R}, E)$ analysis also reveals that the internal electronic structure of representative $4a_0 \times 4a_0$ 'tiles' consists of $3\times 3$ intense incommensurate conductance maxima, whose spectral weight is related to the strong $\mathbf{q}=(\pm 3/4,0); (0, \pm 3/4)2\pi/a_0$ peaks in $g(\mathbf{q}, E)$.

The electronic structure within a representative $4a_0 \times 4a_0$ 'tile' is examined directly in Fig. 3b, which shows $g(\mathbf{r}, E=24\text{ meV})$ measured within the blue box in Fig. 1e. Comparison between Fig. 3b and the simultaneously acquired topography Fig. 3c shows that the lowest conductance coincides with the perimeter atoms of a $4a_0 \times 4a_0$ 'tile', but that the local conductance maxima are not associated with atomic locations inside the tile (except for the central atom). Figure 3d shows the topographic signal (black) and simultaneously measured conductance (red) along a line through the tile centre.

The energetic changes occurring within a representative 'tile' are even more complex. In Fig. 3e, we show averages of spectra selected from representative regions of low conductance in Fig. 3b (white dots in Fig. 3c) and high conductance (red dots in Fig. 3c—see also Supplementary Fig. 3). From this we see that the pseudogap magnitude varies on the atomic scale by a factor of up to three. High pseudogap regions have low subgap conductance but also support a strong resonance at $\Omega=150$ meV (arrow in Fig. 3e).

Next we consider the doping dependence of these phenomena. Bulk electronic transport studies at low temperatures show $Ca_{2-x}Na_xCuO_2Cl_2$ to be insulating at $x=0.06$, marginally conducting at $x=0.08$, and superconducting for $x\geq 0.10$ (Supplementary Fig. 4). The doping dependence of our subgap $g(\mathbf{r}, E)$ is summarized by Fig. 4. The diminishing bias-voltage asymmetry in conductance spectra with increasing $x$ (Supplementary Fig. 1) indicates that top $CuO_2$ plane doping increases as expected. On the other hand, the basic $4a_0 \times 4a_0$ checkerboard state with its $4/3a_0 \times 4/3a_0$ modulations and pseudogap variations changes subtly so that the $A(\mathbf{R}, E)$ (insets in Fig. 4) appear very similar between dopings. Thus, the checkerboard electronic state appears for $x \leq 0.08$ in the ZTPG regime but continues to exist for $p > p_{SC}$, the critical doping for superconductivity.

We recently introduced a conjecture on underdoped Bi-2212 electronic structure[4,14] which can help to clarify these observations. In $\mathbf{k}$-space we distinguish two regions. The first consists of low-energy states on the Fermi arc surrounding the gap nodes. They support a nodal metal at $p < p_{SC}$[9], which becomes a superconductor at $p > p_{SC}$. Second, the high-energy states near the first Brillouin zone-face are incoherent at low doping owing to localization in an electronic crystal state[4]. At low dopings, these high-energy electronic crystal states are unperturbed by the conversion of low-energy states from a nodal metal to a superconductor. This conjecture predicts that the Na-CCOC checkerboard state would be unperturbed by the transition to superconductivity for dopings near $p_{SC}$, consistent with observations.

Angle-resolved photoemission spectroscopy (ARPES) studies of Na-CCOC do indeed detect two very different $\mathbf{k}$-space regions, a large (~200 meV) pseudogap around $(\pi, 0)$, plus a coexisting nodal metal or superconductor consisting of states with $E < 10$ meV[9] near the nodes—all consistent with our conjecture. In Bi-2212 the near-nodal states at $E < 35$ meV are identifiable with superconductivity via quasiparticle interference[4]. But in $x=0.12$ Na-CCOC, the near-nodal states occur for $E < 10$ meV (ref. 9 and Supplementary Fig. 5), a voltage range where we do not yet have enough sensitivity to detect interference patterns (in any material[4]).

The relationship between the checkerboard electronic crystal state in Na-CCOC and the novel electronic phases proposed for copper oxides[14–30] is indeterminate. In the absence of disorder, the orbital-current ordered states[15–17] seem inconsistent with the observations. Although elementary one-dimensional stripes[18–22] also appear inconsistent because of the 90° rotational symmetry of $g(\mathbf{q}, E)$, defect-dominated stripe states[23] or nematic liquid crystal stripes[24] may still be consistent. Checkerboard states due to Fermi surface nesting[14], valence-bond solids[22,25], pair density waves[26,27], hole-pair[22,27,28] or single hole[14,22] Wigner crystals, or the more recently discussed electronic 'supersolids'[29,30], are not inconsistent with data herein. However, none of these proposals captures the intricate spatial and electronic characteristics found in $Ca_{2-x}Na_xCuO_2Cl_2$.

Detailed theory of atomic-scale spectroscopic signatures for the various phases will be required to discriminate between them.

Independently of its microscopic identity, the Na-CCOC checkerboard state is significant because it exemplifies a (no longer hidden) electronic order associated with the copper-oxide pseudogap, and this order is not a translationally invariant liquid of electronic excitations but rather some form of electronic crystal.

Supplementary Information accompanies the paper on www.nature.com/nature.

**Acknowledgements** We acknowledge and thank P. Coleman, E. Demler, M. Franz, J. E. Hoffman, P. A. Lee, K. Machida, K. McElroy, D. Pines, S. Sachdev, T. Senthil, T. Timusk, M. Vojta and J. Zaanen for discussions and communications. This work was supported by the ONR, NSF, MEXT, JST and NEDO. C.L. acknowledges support from a NSERC Postdoctoral Fellowship and Y.K. from a JPSJ Research Fellowship for Young Scientists.

**Competing interests statement** The authors declare that they have no competing financial interests.

**Correspondence** and requests for materials should be addressed to J.C.D. (jcdavis@ccmr.cornell.edu) or T.H. (hanaguri@riken.jp).

**Figure 1** Atomic-scale explorations of electronic states in $Ca_{2-x}Na_xCuO_2Cl_2$. **a**, A schematic phase diagram of hole-doped copper-oxides showing the Mott insulator, high-$T_c$ superconductor (HTSC) and metallic phases along with the 'pseudogap' regime and the ZTPG line. **b**, Crystal structure of $Ca_{2-x}Na_xCuO_2Cl_2$. Red, orange, blue and green spheres represent Ca(Na), Cu, O and Cl atoms, respectively. Conducting $CuO_2$ planes are sandwiched by insulating CaCl layers. **c**, A characteristic spatially averaged tunnelling

conductance spectrum of $x$=0.12 Na-CCOC. The large particle–hole asymmetry in conductance at high energies can be associated with the light doping of a Mott insulator (see text). At low energies a skewed V-shaped gap exists. At energies below ~10 meV, changes occur in the spectra of superconducting samples (see Supplementary Fig. 5). The spectrum measured on equivalently underdoped Bi-2212 is shown in blue. **d**, High-resolution STM topograph of the cleaved CaCl plane of a crystal with $x$=0.10. The perfect square lattice, without discernible bulk or surface crystal reconstructions, is seen. The image was taken at a junction resistance of 4 GΩ and sample bias voltage $V_s$=+200 mV. **e**, The conductance map $g(\mathbf{r}, E)$ at $E$=+24 meV in the field of view of **d**. It reveals strong modulations with a $4a_0 \times 4a_0$ commensurate periodicity plus equally intense modulations at $4a_0/3 \times 4a_0/3$ and strong modulations at $a_0 \times a_0$. All data in this paper are acquired near $T$=100 mK in a dilution refrigerator-based scanning tunnelling microscope.

**Figure 2** Energy dependence of the modulation wavevectors in the electronic crystal state. **a–c**, Images of differential conductance $g(\mathbf{r}, E)$ at $E$=+8, +24 and +48 meV in the same field of view as Fig. 1d; junction resistance was set to 2 GΩ at $V_s$=+200 mV. Such 'checkerboard' patterns are observed at all energies $|E|$<100 meV with strong correlations between the patterns at each energy. The dark low-conductance one-dimensional streaks are always registered to the atomic rows. The modulations are intense, using up to 90% peak-to-peak of the mean conductance at each energy. **d–f**, The calculated $g(\mathbf{q}, E)$ are Fourier transforms of the $g(\mathbf{r}, E)$ in **a–c**. Regions near the centre of each $g(\mathbf{q}, E)$ reflect the nanoscale electronic disorder. The modulations in $g(\mathbf{q}, E)$ contain three different **q**-vectors: $\mathbf{q}=(\pm 1, 0); (0, \pm 1)2\pi/a_0$, $\mathbf{q}=(\pm 1/4, 0); (0, \pm 1/4)2\pi/a_0$, and $\mathbf{q}=(\pm 3/4, 0); (0, \pm 3/4)2\pi/a_0$. Fourier transforms of these $g(\mathbf{r}, E)$ functions do not break 90° rotational symmetry, within the systematic uncertainty of the measurement as determined by the degree to which the atomic modulations $\mathbf{q}=(\pm 1, 0); (0, \pm 1)2\pi/a_0$ break this symmetry. All three dominant **q**-vectors do not disperse with energy. **g–i**, The magnitude of the Fourier transform is plotted along the $(2\pi, 0)$ (black filled circles) and $(0, 2\pi)$ (red open circles) directions for each of these same $g(\mathbf{r}, E)$ and $g(\mathbf{q}, E)$ as in **a–c** and **d–f**.

**Figure 3** Electronic structure imaging within a representative $4a_0 \times 4a_0$ 'tile'. **a**, Autocorrelation image of typical $|E|$<100 meV LDOS maps showing the $4a_0 \times 4a_0$ structure inside a dark perimeter, which is exactly at the locations of the 16 atoms at the common perimeter between adjacent $4a_0 \times 4a_0$ regions. **b**, A representative tile of the $4a_0 \times 4a_0$ state is seen directly. The tile exhibits a very low conductance at the perimeter and high LDOS conductance pattern with nine incommensurate maxima inside. The average spatial electronic structure in **a** is remarkably consistent with electronic structure

of this (and other) tiles, but this should not be overemphasized because there is also a great deal of variability. Junction resistance for measurement was 2 GΩ at $V_s$=+400 mV. **c**, The simultaneously acquired topographic image showing the locations of what are believed to be the Cl atoms (light) above each Cu atom in the $CuO_2$ plane. **d**, The topographic signal (black) and simultaneously measured conductance (red) along the line shown in **c** through the tile centre. **e**, Spectra show V-shaped pseudogaps regardless of the positions. The most remarkable difference between dark (low conductance) and bright (high conductance) spots in **b** appears in the positive bias voltage. At dark spots, a strong peak is observed at $\Omega$≈+150 meV. On average the high pseudogap (most insulating) regions exhibit a strong resonance at $\Omega$≈+150 meV so that the $g(\mathbf{r}, E$=150 meV) is anticorrelated with the $g(\mathbf{r}, E)$ below 100 meV.

**Figure 4** Doping dependence of electronic structure images. The $g(\mathbf{r}, E)$ patterns seen at three different dopings ($x$=0.08, 0.10, 0.12) are highly characteristic of lightly doped Na-CCOC. Although the high-energy spectra are changing (Supplementary Fig. 1), the subgap spatial characteristics are almost the same in the non-superconducting and superconducting phases in the energy range 10 meV<|$E$|<100 meV. The autocorrelation functions seen as insets appear very similar except for a slight increase in correlation length with increasing doping. The white scale bar in the inset corresponds to 2 nm.

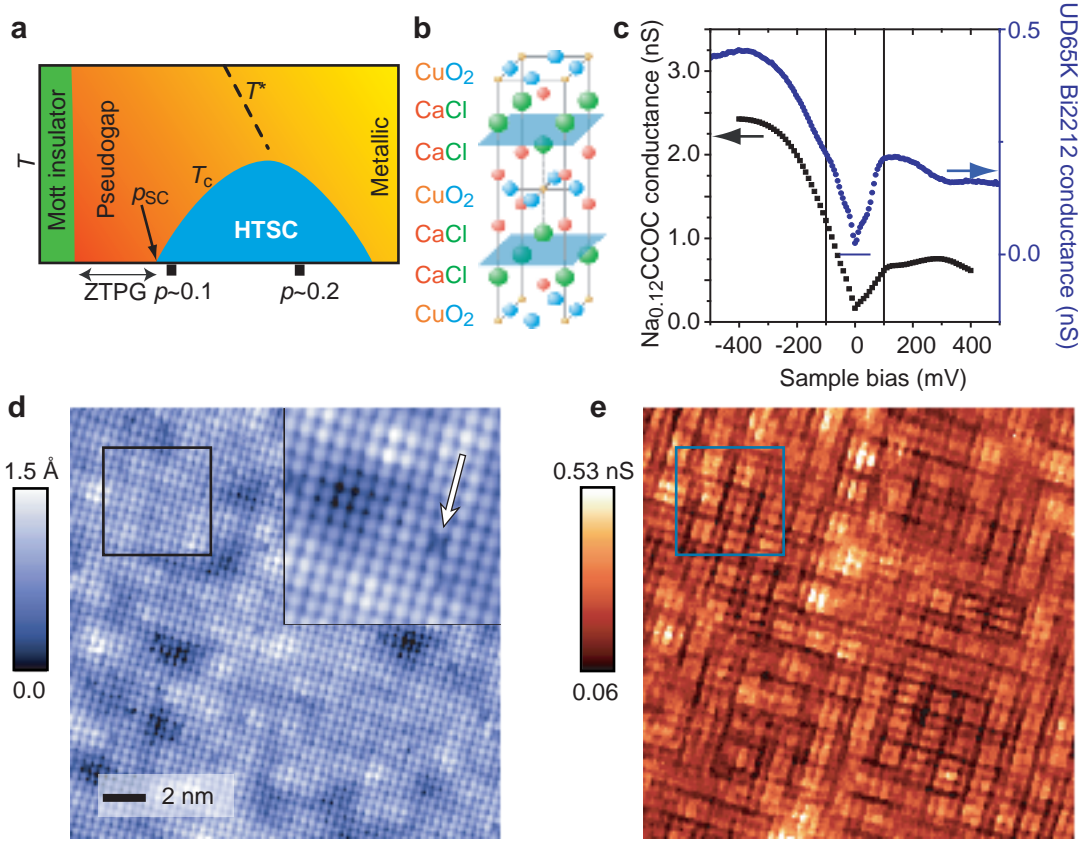

**Fig. 1**

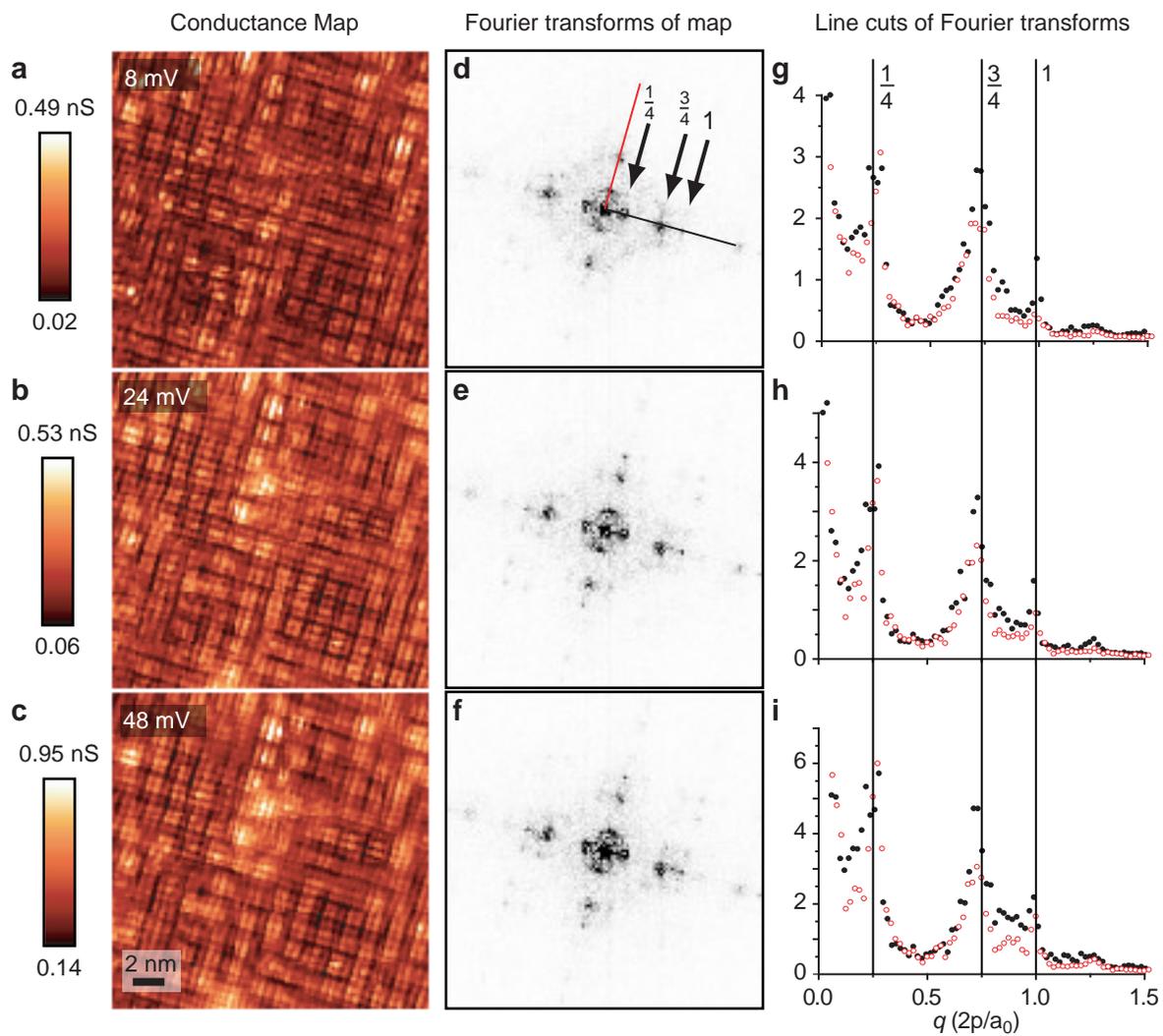

**Fig. 2**

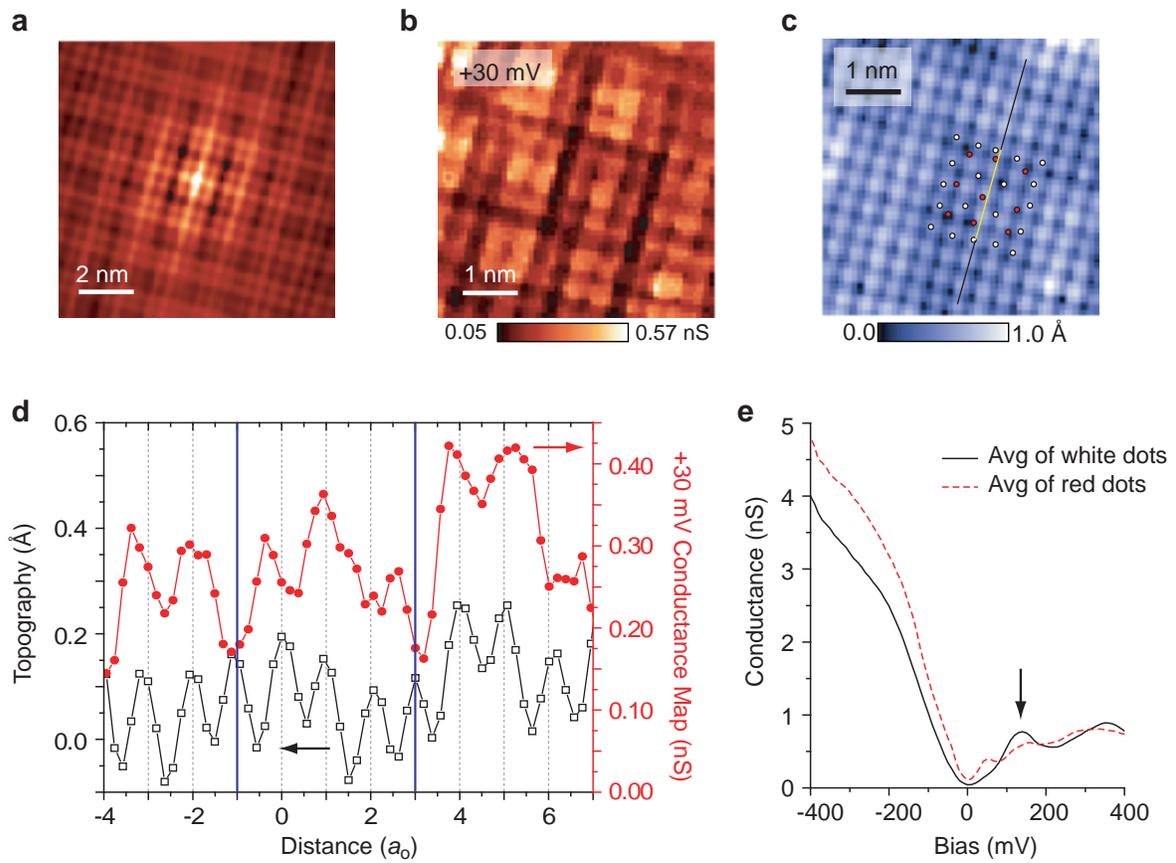

**Fig. 3**

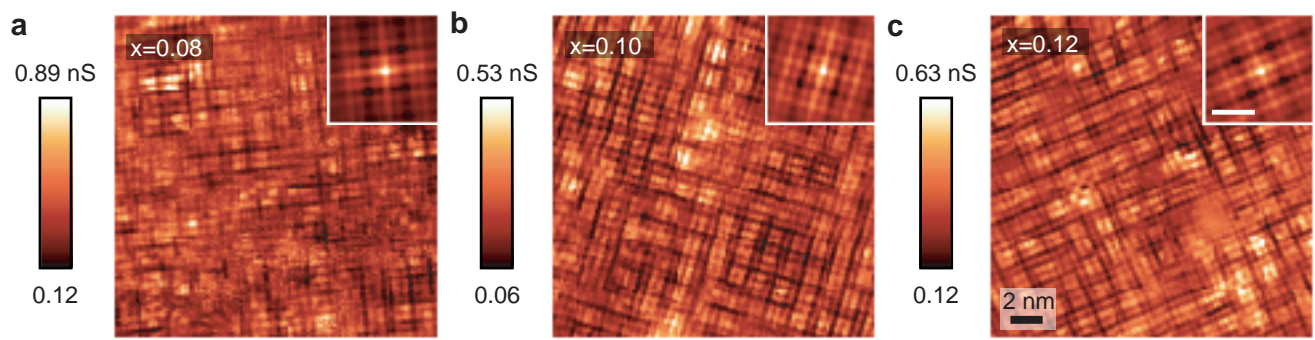

**Fig. 4**